\documentclass[12pt]{iopart}
\usepackage{iopams}
\usepackage{graphics}

\newcommand{\imag}{\mathop{\mathrm{Im}}}
\newcommand{\ch}{\mathop{\mathrm{ch}}}
\newcommand{\EF}{\mathcal{E}}
\newcommand{\Ut}{U_{\tau}}
\newcommand{\Ft}{F_{\tau}}
\newcommand{\pst}{\psi_{\tau}}

\newcommand{\pF}{\psi^{\mathrm{F}}}
\newcommand{\pFt}{\psi^{\mathrm{F}}_\tau}
\newcommand{\pFtn}[1]{\psi^{\mathrm{F}}_{#1}}
\newcommand{\PF}{\mathcal{P}}
\newcommand{\abs}[1]{\left\vert #1 \right\vert}
\newcommand{\norm}[1]{\left\Vert #1 \right\Vert}

\newcommand{\ket}[1]{\vert #1 \rangle}
\newcommand{\brkt}[2]{\langle #1 \vert #2 \rangle}
\newcommand{\braket}[3]{\left\langle #1\left\vert\,#2\,\right\vert
#3\right\rangle}
\newcommand{\ketbra}[2]{\vert #1 \rangle\langle #2 \vert}

\newcommand{\schrod}{Schr\"odinger}
\newcommand{\proof}{{\bf\em Proof.}}
\newcommand{\qed}{\hspace{\stretch{1}} $\blacksquare$}
\newtheorem{thm}{Theorem}
\newtheorem{lemma}[thm]{Lemma}
\newtheorem{prop}[thm]{Proposition}
\newtheorem{cor}[thm]{Corollary}
\newcommand{\msize}[1]{$#1 \!\times\! #1$}

\begin{document}

\title{Quantum Response at Finite Fields and Breakdown of Chern Numbers}
\author{J E Avron and Z Kons
\footnote[1]{E-mail addresses: {\tt avron@physics.technion.ac.il}
and {\tt konsz@physics.technion.ac.il}}}

\address{Department of Physics, Technion, 32000 Haifa, Israel.}

\begin{abstract}
We  show that the response to an electric field, in  models of the
Integral Quantum Hall effect, is analytic in the field and has
{\em isolated essential singularity} at zero field. We also study
the breakdown of Chern numbers associated with the response of
Floquet states. We argue, and give evidence, that the breakdown of
Chern numbers in Floquet states is a discontinuous transition at
zero field. This follows from an observation, of independent
interest, that Chern numbers for finite dimensional Floquet
operators are generically zero. These results rule out the
possibility that the breakdown of the Hall conductance is a phase
transition at finite fields
 for a large class of models.
\end{abstract}

\section{Introduction}
The principal motivation for the present work is the question: Is
the breakdown of the Integer Quantum Hall effect a (quantum) phase
transition? Since the Hall conductance  in the adiabatic limit is
identified with a Chern number, the question can also be phrased
as: Is the breakdown of Chern numbers a phase transition?

Experimentally,
\cite{KlitzingBRK,CageBRK1,CageBRK2,Kawaji1,Kawaji2,komiyama,rigal,watts}
the breakdown of the Hall effect at finite driving currents  is
signaled by the onset of dissipation, and is accompanied by
hysteresis and complex dynamical behavior. The critical current
and voltage depend, in general, on the geometry of the system, the
temperature and on the magnetic field. It has been suggested that
a phase diagram for the breakdown resembles the phase diagram of
superfluidity \cite{rigal}.

Is there a theoretical basis for identifying breakdown with a
phase transition? Naively, one can argue both ways.  In a  class
of models of the Integer Hall effect  the Hall conductance (at
zero temperature and in the limit of linear response) is related
to a Chern number \cite{ThoulessBOOK}.  Chern numbers, being
integers,  depend discontinuously, if at all, on parameters in the
Hamiltonian. So, if the strength of the external electric field
was just like any other parameters in the Hamiltonian, one would
expect the breakdown of the Hall conductance to be discontinuous
and at non-zero field. This would say that the breakdown of the
Hall effect is indeed a quantum phase transition at finite fields.

On closer inspection one realizes that this line of reasoning can
not be quite right. For, if it was, then the Hall conductance would
remain precisely quantized also for small but finite values of the
electric field. This would imply that there are no corrections to
the quantization -- not even exponentially small correction.
Common wisdom is that while there are no power corrections to  the
integral Hall conductance, there are exponentially small
corrections \cite{Seiler}. As we shall explain in detail, and this
is going to be a key point of our analysis, the strength of the
driving electric field (or the driving emf) is a special parameter
which affects the Hamiltonian and the  evolution in a way that is
structurally different from say, the strength of the magnetic
field or the disorder potential.

A simple and common argument, with some experimental support, says
that the breakdown occurs at finite driving fields so that the
critical field, $\EF_c$, scales like $B^{3/2}$, where $B$ is the
strength of the magnetic field. This estimate follows from comparison of
the energy gap in Landau levels with the voltage drop on a
magnetic length. The breakdown is then attributed to tunneling
between Landau levels. This argument does not directly address the
question if the breakdown is a phase transition. It also has a
weakness  in that the Landau Hamiltonian  with constant electric
and magnetic fields is explicitly soluble and does not show
breakdown.   For other theories of the breakdown see e.g.\
\cite{ThoulessBRK} and references therein.

To address the breakdown as a quantum phase transition, a handle
on the conductance at finite  fields  is needed. This goes beyond
linear response and Kubo's formulation.

In this work we shall concentrate on the breakdown of  the Hall conductance,
rather than the breakdown that occurs in the dissipative conductance in the Hall
effect. This is done for two reasons. The first is that we are interested in
breakdown
that occurs in Chern numbers, for which the Hall conductance is a basic paradigm.
The
second is for concreteness sake. Some parts of what we say can be
transcribed, {\it mutatis mutandis}, for the dissipative conductance.

We find no theoretical support to the hypothesis that the
breakdown in the Hall effect is a phase transition at {\em finite
fields}. Rather, we find support to the claim that the conductance
has an {\em essential singularity at zero fields}, which can, of
course, manifest itself in what resembles a phase transition.

In section \ref{sec:scale} we show that the strength of the
driving electric field (or emf), $\EF$, can be related to a time
scale $\tau$. In section \ref{sec:analyticity}  we show that the
expectation value of a (bounded) observable, in particular, the
current density in tight-binding models of non-interacting
electrons, is an analytic function of $\EF$ with an essential
singularity at $\EF=0$ when
 $\EF t$ is kept fixed.
In section \ref{sec:chern} we consider the analytic properties of
a natural notion of transport for a class of model Hamiltonians
which, in the limit of linear response, reduces to the usual
notion of conductance that coincides with  a Chern number. We show
that this observable is analytic in $\EF$ with an essential
singularity at $\EF=0$.  It is interesting that the absence of
phase transition at finite fields can be shown for the same class
of models where one can prove quantization.
In section \ref{sec:harper} we study the Harper model for which we
present  numerical results.
In section \ref{sec:floquet} we study  Chern numbers associated with
Floquet states, their properties and their interpretation as Hall
conductance of the Floquet states.
In section \ref{sec:breakfloquet} we study the breakdown of Chern
numbers associated to Floquet states. We show that the breakdown is
discontinuous and argue that it occurs at zero fields, $\EF=0$.
In section \ref{sec:numeric} we describe  numerical evidence that
supports the claim that non-zero Chern numbers for Floquet states
are unstable against perturbations in the Hamiltonian.

\section{Driving Fields Interpreted as a Time Scale}\label{sec:scale}

Because gauge invariance allows one to impose one condition on the
scalar and vector potential, it is always possible to choose a
gauge where the {\em external} electric and magnetic fields are
described only by the vector potential $\vec A$. In order to
produce electric field this potential has to be time dependent.
Assume that this dependence is characterized by some time scale
$\tau$ as $\vec A(\vec x,t/\tau)$.
Suppose now that we scale time so that $t=\tau\, s$. Maxwell
equation gives the external electric field, (in scaled time), as
\begin{equation}\label{eq:maxwell}
\vec E(\vec x,s)= -\frac{1}{\tau c} \, \partial_s {\vec A}(\vec
x,s)\:.
\end{equation}
With $\vec A$ fixed, weak electric fields correspond to large
$\tau$ while strong electric fields correspond to small $\tau$. In
systems that are otherwise time independent, $\tau$ interpolates
between weak and strong fields. A similar argument can be made
about the emf, which is a line integral of the electric field.

The identification of the time scale $\tau$ with the strength of
the external driving, be it  the electric field or an emf, is not
a new idea, of course. It lies at the heart of the identification
of the adiabatic limit with linear response. What is perhaps new
here is that we want to use this correspondence for any driving,
and in particular identify strong driving fields and large emfs
with short time scales. Equation \eref{eq:maxwell} suggest that we
write
\begin{equation}\label{eq:EFasTau}
\EF= \left(\frac{\hbar}{e }\right)\,\frac1{\tau} \:,
\end{equation}
where $\EF$ is a measure of the strength of the driving field. We
have put fundamental constants into this relation so that (in cgs
units) $\EF$ has the dimensions of an emf (or voltage). Because
lattice models also have a natural length scale -- the lattice
spacing, one can choose constants so that $\EF$ has dimensions of
an electric field. The identification of the strength of the field
with an inverse time scale is central to our considerations and
turns out to have consequences for transport. The first and easy
consequence is that the question of phase transition in, say, the
Hall conductance as function of the driving electric field, can be
phrased as a question about the analytic properties of the Hall
conductance as a function of $\tau$. The second consequence is
discussed in the next section.

\section{Analyticity of Observables in $\tau$}\label{sec:analyticity}
The identification of the time scale with the strength of the
driving makes the driving field, be it an electric filed or an
emf, a special parameter in the Hamiltonian. This can be seen from
the form of the \schrod\ equation for the time evolution operator
in scaled time:
\begin{equation}\label{eq:schrodinger}
 \rmi\, \dot \Ut(s) = \tau H(\vec A(s) ) \Ut(s) \:, \quad
\Ut(0)=1 \:.
\end{equation}
The $\tau$ dependence of the Hamiltonian  is linear, and
a-fortiori, analytic in $\tau$, irrespective of how $H$ depends on
$\vec A$. This has the consequences:
\begin{prop}\label{prop:analy}
Suppose that $H(\vec A(s))$ is  bounded and self-adjoint, then $\Ut(s)$ and
$\Ut^\dagger (s)$ are both  entire functions of $\tau$.
\end{prop}
\proof\ The first part follows from a standard argument about the
absolute convergence of the Dyson series for $\Ut$.
The second assertion is  a consequence of the fact that, for real $\tau$,
$\Ut^\dagger$ satisfies
\begin{equation}\rmi\, \dot \Ut^\dagger (s) =
 -\tau \,\Ut^\dagger (s)H(\vec A(s) )  \:, \quad
\Ut^\dagger(0)=1 \:.
\end{equation}
The Dyson series implies that one can extend  $\Ut^\dagger (s)$ to
an entire function of $\tau$. \qed\newline
 It follows that:
\begin{cor}\label{cor:analy}
Suppose that $I$ is a (fixed) bounded operator, and
 $\psi$ a  ($\tau$ independent) initial state then:
\begin{equation}
I(\tau,s)= \braket{\psi}{\Ut^\dagger (s)\, I \, \Ut(s)}{\psi}
\end{equation}
is an entire function of $\tau$.
\end{cor}
This leads to the main result of this section:
\begin{thm}
Let $H(\vec A )$ be bounded self-adjoint operator.  Let $I$ be a
bounded observable, then, its expectation value $\langle I\rangle
(\EF,s)$ is an analytic function of $\EF$ with  isolated essential
singularity at $\EF=0$ and with Laurent expansion
\begin{equation}\label{eq:Laurent}
\langle I\rangle (\EF,s)=\sum_{n=0}^\infty \frac{a_n(s)}{\EF^n} \:,
\end{equation}
with infinitely many of the $a_{n}\neq 0$ (except if $\langle
I\rangle$ is a constant).
\end{thm}
\proof\ From proposition \ref{prop:analy}, $\langle I\rangle
(\EF,s)$ is an analytic function whose Taylor expansion about
$\tau=0$ is absolutely convergent with radius of convergence that
is infinitely large.  The expansion can not have positive powers
of $\EF$, for if it did, the response at $\tau=0$ would not be
analytic. The singularity at the $\EF=0$ must be essential for if
was a pole then the response would diverge as $\tau\to\infty$
along any direction. But, the response is  bounded on the real
$\tau$ axis by the self-adjointness of the Hamiltonian. It follows
that the response can not have a pole of finite order at $\EF=0$.
\qed

Remarks:
\begin{enumerate}
\item It is important for the conclusion of the theorem to hold
that the scaled time $s$ is kept finite. If one lets $s\to\infty$
then, analyticity in $\tau$ may be lost.

\item We have restricted ourselves to bounded Hamiltonians and bounded
observables. This is because phase transitions are normally a long
wavelength, low energy phenomenon. Some parts of the theorem can also be
extended to unbounded \schrod\ type Hamiltonians provided  some care is taken
about questions of domains of operators.

\item The smooth dependence of the evolution in $\tau$ is not
really special to quantum mechanics.  It holds also in classical
mechanics under slightly stronger conditions, namely that
$\partial_p H$ and $\partial_x H$ are both bounded functions
(provided the initial state is analytic in $\tau$).

\item A classical model that shows breakdown of analyticity in a
constant electric field is the washboard potential with initial
state  at rest at a local minimum. When $\EF$ is sufficiently small,
the velocity stays zero for all times. When $\EF$  passes a threshold the
particle accelerates indefinitely. This is not a
counterexample   to the analyticity of observabales
because the initial state in the washboard potential is $\EF$
dependent in a non-analytic way.

\item Two prototype functions for $\langle I \rangle$ that
 satisfy the conclusion
of the theorem are $\exp(-c/\EF^2) $ and $\sin(c/\EF^2)$.

\item The theorem has an interesting implication to the question of
power law corrections to linear response. It follows from the
theorem that if $\langle I\rangle$ has an asymptotic expansion in
powers of $\EF\in \mathbb{R}_+$ at $\EF=0$ then this expansion
must vanish identically. One may wonder if the absence of power
law corrections to linear response is  a valid conclusion of
equation~\eref{eq:Laurent}. It is not, as one can see from the
second prototype example in (v), where an asymptotic expansion in
powers of $\EF\in \mathbb{R}_+$ does not exist.

\item Absence of power corrections to linear response in
quantized Hall conductance
has been proven rigorously for the quantum Hall effect by Klein
and Seiler \cite{Seiler}. Their proof uses an adiabatic theorem to
all orders, to show that an  asymptotic expansion exists and a
clever trick  that shows that the coefficients must vanish. The
theorem above can be used to replace their clever trick.
\end{enumerate}

In tight-binding models the current operator, for finitely many
interacting electrons or the current density operator for
infinitely many non-interacting electrons, is a bounded. It
follows that the currents in such models have an essential
singularity at $\EF=0$, but are analytic at non-zero $\EF$
provided the time $t\EF$ is kept fixed.

\section{Breakdown of Chern Numbers}\label{sec:chern}

In this section we consider the breakdown of an observable
associated with quantized charge transport. Quantized charge
transport  occurs for  a class of Hamiltonians in the adiabatic
limit. This class of Hamiltonians includes models of the quantum
Hall effect, and in particular includes the Harper model. It also
includes certain models of the Hall effect with electron-electron
interactions. The observable that we consider reduces to the Hall
conductance in the limit of linear response, and coincides with a
Chern number. In  this section we discuss the analytic properties
of this observable with $\tau$.

The model Hamiltonians  for which quantization occurs have the
following struc\-ture \cite{Stone,ThoulessBOOK}: $H(\phi,k)$ is a self-adjoint
Hamiltonian that depend periodically on two real, dimensionless
parameters, $\phi$ and $k$, with period $2\pi$. $H(\phi,k)$ may be
associated with a {\em finite multiparticle} system,  where $\phi$
and $k$ are external parameters, for example, two Aharonov-Bohm fluxes.
 Alternatively, $H(\phi,k)$ may be a Bloch type
Hamiltonian in two dimensions describing {\em infinitely many
non-interacting} electrons, where $\phi$ and $k$ are two Bloch
momenta.   In either case, we shall require that for fixed $\phi$
and $k$ the Hamiltonian $H(\phi,k)$ has discrete spectrum with no
eigenvalue crossing. For the sake of simplicity we assume that the
$\phi$ and $k$ dependence is smooth and that $H(\phi,k)$ is a
bounded operator such as a tight binding model and its
multiparticle generalizations.

The time dependence comes from a time dependence of $\phi$ on a
time scale $\tau$\footnote{Here $\phi$ plays the role of $\vec A$
of the previous section.}. We suppose that $\phi(s)$ is a smooth,
monotonically non-decreasing function of $s$ with $\phi(s)=0$ in
the past, $s<0$, and $\phi(s)=2\pi$, in the future, $s\ge 2\pi$.
$H$ is therefore time dependent only on a finite interval of
(scaled) time $[0,2\pi]$.

Since $ \partial_{k} H $ is the current operator in these models
the total charge transported by the action of $\phi$ is:
\begin{equation}\fl
Q(\tau;\psi)= \frac{\tau}{2\pi}\int_{0}^{2\pi}\rmd k
\,\int_0^{\infty} \rmd s \braket{\Ut(s,k)\psi(k)}{ \frac{\partial
H(\phi(s),k)}{\partial k }\,}{\Ut(s,k)\psi(k)} \:,
\end{equation}
with initial condition $\psi$ that is an eigenstate of $H(\phi,k)$
for $s=0$.

There are several special things that happen in the adiabatic
limit. First,  $Q$ coincides with the Hall conductance defined via
Kubo's formula \cite{Stone,ThoulessBOOK}.  Second, it is
independent of the functional form of $\phi$ (provided $\phi$
satisfies the limiting conditions). Third, since in the adiabatic
limit there is no current once the driving stops, i.e. when
$\dot\phi=0$,  $Q$ can also be written as
\begin{equation}\label{eq:Q}
Q(s;\tau,\psi)= \frac{\tau}{2\pi}\int_{0}^{2\pi}\rmd k
\,\int_0^{s} \rmd s' \braket{\Ut\psi}{ \frac{\partial
H}{\partial k }\,}{\Ut\psi} \:,
\end{equation}
provided $s\ge 2\pi$. $Q$ is always a measure of the charge
transport, but its identification with a conductance is valid, in
general, only in the adiabatic limit\footnote{ Equation
\eref{eq:Q} implies that $Q$ vanishes in the limit $\tau\to 0$,
i.e. in the limit of {\em large} external fields, contrary to
common experience. One reason for this is that  tight binding
models are unreasonable when the external fields are large on
atomic scale. We shall take the point of view that the breakdown
is a low field phenomenon that is divorced from the asymptotic
behavior at very large fields.}.

Applying the results of the previous section we see that the
breakdown of the Chern number $Q$ is smooth, and has no phase
transition at finite fields. Due to the prefactor $\tau$ in
equation \eref{eq:Q}, $a_0=0$ in equation \eref{eq:Laurent}.

It is interesting that for the class of models where one can
prove that $Q$ is quantized in the adiabatic limit, one can also show that
it is an analytic function of the field away from $\EF=0$. (For model
Hamiltonians with infinitely many interacting electrons there is, at
present, no proof of quantization either.)

There are now three possibilities. The first, and perhaps
simplest, is that the breakdown of the Hall effect is  a
consequence of an essential singularity at zero field. The second
is that the breakdown of the Hall effect is associated with the
limit $s\to\infty$. And the third is that the breakdown is a
property of infinitely many interacting electrons. We shall
examine the second possibility in section \ref{sec:floquet}.

\section{Example: Hall Conductance in the Harper Model
}\label{sec:harper}

The Harper model is the simplest, non-trivial, model where one can
study the breakdown of the Hall effect in detail, at least
numerically. The model  is associated with a square lattice,
$\mathbb{Z}^2$; an external {\em homogeneous} magnetic field $B$,
and {\em homogeneous} electric field $\EF$ pointing in the $x$
direction. We choose a gauge so that the electric field is
described by a time dependent vector potential. After separation
of variable, the model is described by a Hamiltonian on
$\mathbb{Z}$ parameterized by one Bloch momentum, $k$.  The
Hamiltonian action on the vector $\Psi\in\ell^2(\mathbb{Z})$,
while the electric field is acting, is:
\begin{equation}\label{eq:TimeHarper1}
 \rme^{\rmi \EF t}\Psi_{x+1}+\rme^{-\rmi\EF t}\Psi_{x-1}  +2\cos(B
 x+k) \Psi_x \:,\quad
 k\in[-\pi,\pi] \:,\quad
 x\in \mathbb{Z} \:.
\end{equation}
For rational magnetic field $B=2\pi p/q$ with $p,q \in \mathbb{Z}$
the Hamiltonian is periodic in $x$, with period $q$. One then
classifies the solutions by a second Bloch momentum,
$\ell\in[-\pi,\pi]$ so that $\Psi_{x+q}=\exp(-\rmi \ell)\Psi_x$.
Fixing periodic boundary conditions  is  achieved by the unitary
transformation $\Psi_x\rightarrow \rme^{\rmi \frac{\ell}{q} x }
\Psi_x$. So, finally, the requisite form of the Harper Hamiltonian
we shall study is:
\begin{eqnarray}\label{eq:TimeHarper2}
 \Big(H(\phi,k)\Psi\Big)_x = \rme^{\rmi \phi }\Psi_{x+1}+\rme^{-\rmi
  \phi }\Psi_{x-1} +2\cos(2\pi \frac{p}{q} x+k ) \Psi_x \:, \nonumber \\
\Psi_{x+q}=\Psi_x ,\quad  \phi(t)= \cases{\frac{\ell}{q},&if
$t<0$;\cr \EF t+\frac{\ell}{q}, &if $0<t<\frac{2\pi}{\EF}$;\cr
2\pi+\frac{\ell}{q}, &otherwise.\cr}
\end{eqnarray}
This corresponds to a \msize{q} hermitian matrix, periodic in
$\phi$ and $k$. The driving electric field, $\EF$, is related to
the adiabaticity parameter $\tau$, by \eref{eq:EFasTau} (in  units
where $e=\hbar=1$). In this example $\phi(s)$ is continuous and
piecewise linear. As a consequence the electric field is
discontinuous in time. It is easy to modify the model so that the
electric field is switched on and off continuously.  We have
examined also such models and the behavior is qualitatively
similar to models with discontinuous switching.

\begin{figure}[htb]
  \centering
  \includegraphics{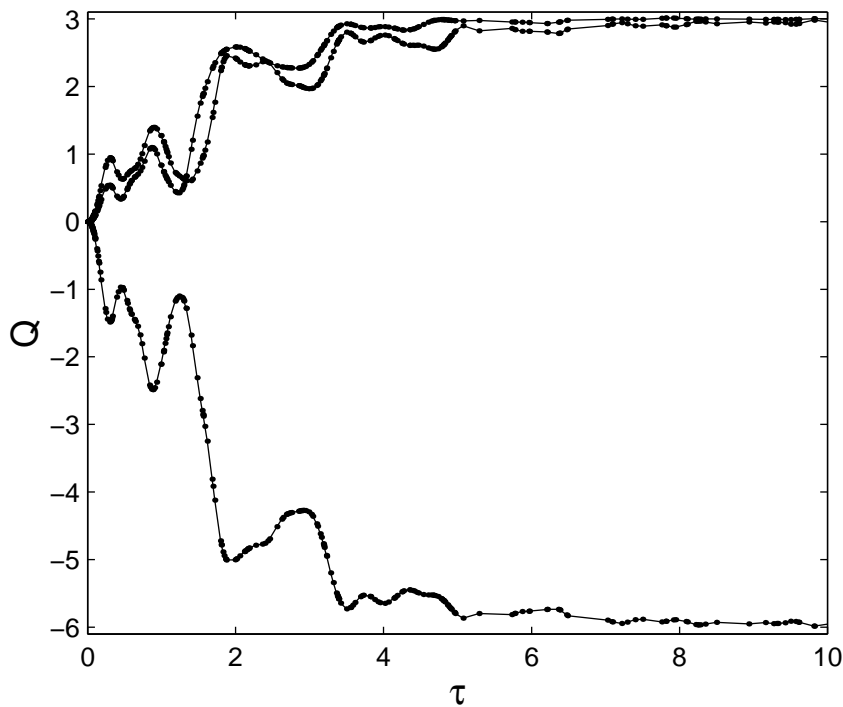}
  \caption{The Hall conductance of the \msize{3} Harper Hamiltonian as
  function of the adiabaticity parameter $\tau$ }
  \label{fig:QH3x3}
\end{figure}

For $p=1$, $q=3$, the Harper Hamiltonian is a \msize{3} matrix.
Its Chern numbers are $\{3,3,-6\}$. \Fref{fig:QH3x3} shows the
charge transport, $Q(2\pi;\tau,\psi)$, as function of the applied
field. The graph has a rich and complex structure, but no sharp
breaking. Substantial deviation from integral quantization occur
near $\tau=5$.


\section{Long-time limit and Floquet states}\label{sec:long}
In this section we  consider the long-time limit of observables,
in time-periodic (finite dimensional) Hamiltonians. An example is
the current density operator  in tight-binding models, such as the
Harper model, driven by a time-independent electric field. One
advantage of using a time dependent representation over the time
independent representation is that one has to deal with finite
dimensional matrices. In the time independent representation the
matrices are infinite dimensional.

In the theory of time-periodic Hamiltonians Floquet states
\cite{cycon,nimrod,Floquet} play a role analogous to that of
eigenstates for time independent operators. We shall see that the
long-time (Abelian) limit of observables in time periodic systems
is related to the expectation value of the observable in Floquet
states.

Consider a time-periodic, self-adjoint, finite dimensional matrix
Hamiltonian $H(s+2\pi)=H(s)$. The Harper model of section
\ref{sec:harper} is an example except that now the electric field
is time independent for all (positive) times $s\ge 0$.

Let $F$ denote the unitary evolution over one
cycle
\begin{equation}
 F=U(2\pi)  \:.
\end{equation}
The time evolution for $n\in\mathbb{Z}$ periods is clearly
\begin{equation}
F^{n}=U(2n\pi) \:.
\end{equation}
We assume that $F$ is a finite dimensional, non-degenerate, matrix.
 Its spectral representation is then
\begin{equation}
 F= \sum_n e^{\rmi E_n} \ketbra{\pFtn{n}}{\pFtn{n}}\:.
\end{equation}
$E_n$ are the \emph{quasienergies}, and $\ket{\pFtn{n}}$ the
eigenvectors of $F$.

Consider the observable associated to the bounded operator $I$
(e.g the current density operator in tight binding models). Let
$\psi$ be the initial state of the system at $s=0$, then the
Abelian, long-time, limit of the expectation value of $I$, when
$\psi$ evolves according to the \schrod\ evolution, is:
\begin{eqnarray}\label{eq:Qavg}
 \lim_{M\to\infty}\frac{1}{M}\sum_{j=1}^{M} \langle I(2\pi j) \rangle
 =
 \lim_{M\to\infty}\frac{1}{ M}\sum_{j=1}^{M}
\braket{F^{j}\psi}{ I}{F^{j}\psi} \nonumber \\ = \sum_{l,n}\,
\braket{\pFtn{n}}{ I}{\pFtn{l}}\,\brkt{l}{\psi}\,\brkt{\psi}{n}
\,\left(
\lim_{M\to\infty}\frac{1}{M}\sum_{j=1}^{M}\,e^{\rmi(E_l-E_n)j}
\right)\nonumber \\ =\sum_l  \abs{\brkt{\psi}{\pF_l}}^2
 \braket{\pF_l}{ I}{\pF_l}  \:.
\end{eqnarray}
The long time behavior is a weighted sum of the expectation value
of the operator $I$ in Floquet states, i.e. $ \braket{\pF_l}{
I}{\pF_l} $.

In particular, if one now defines the Hall conductance as the ratio of
the Hall current density to the electric field in the long time limit, then the
breakdown of the Hall effect is related to analytic properties of the
expectation value of the current density operator in Floquet states.
This is the subject of the following sections.

\section{Transport in Floquet States}\label{sec:floquet}
In this section we recall, and extend, results of Ferrari \cite{Ferrari} that
 relate the Hall conductance of Floquet states to
their Chern number and the winding
numbers of their quasienergies.

Consider a time-periodic, self-adjoint, finite dimensional matrix
Hamiltonian $H(s+2\pi,k)=H(s,k)$, which depends analytically on
$s$ and $k$. The Harper model of section \ref{sec:harper} is an
example except that now the electric field is time independent for
all (positive) times. $s$ is, as before, the scaled time.

We define the Floquet operator as the unitary evolution over one
cycle
\begin{equation}\label{eq:FlqDef}
 \Ft(s,k)=\Ut(s+2\pi,k) \Ut^\dagger(s,k) \:.
\end{equation}
Floquet theorem can be expressed as
\begin{equation}
\Ft(s+2\pi,k)=\Ft(s,k) \:.
\end{equation}
We assume that $\Ft$ is a finite dimensional matrix. It therefore
has discrete spectrum and its spectral representation is
\begin{equation}\label{eq:FlqEigens}
 \Ft(s,k)= \sum_n e^{\rmi E_n(s,k;{\tau})} \PF_n(s,k;\tau) \:.
\end{equation}
$E_n$ are the \emph{quasienergies}, and $\PF$ are the
eigenprojections. We shall denote by $\ket{\pFt}$ a unit
eigenvector of $\Ft$.

\begin{lemma}\label{lem:FlqSrd}
For a Floquet operator generated by a bounded and analytic
$H(s,k)$,
\begin{enumerate}
\renewcommand{\theenumi}{\arabic{enumi}}
\item The quasienergies are independent of $s$,
$E_n(s,k;\tau) = E_n(k;\tau)$.

\item The eigenfunctions (eigenprojections) obey the \schrod\
(Heisenberg) equations
\begin{eqnarray}
  \rmi \partial_s \ket{\pFt(s,k)}=\tau H(s,k) \ket{\pFt(s,k)} \:,
  \\
\label{eq:PFSchrod}
  \rmi \partial_s \PF(s,k;\tau)=\tau [H(s,k),\PF(s,k;\tau)] \:.
\end{eqnarray}

\item $\Ft(s,k)$ is an entire function of $\tau$ and analytic function
of $s$ and $k$.
\end{enumerate}
\end{lemma}
\proof\ The Floquet operator satisfies
\begin{equation}\label{eq:Floquet}
 \Ft(s,k) = \Ut(s,k)\Ft(0,k)\Ut^{\dagger}(s,k) \:,
\end{equation}
so its eigenvalues are independent of $s$. The eigenfunctions and
eigenprojections  transform in the same way
\begin{eqnarray}
 \label{eq:Psitrans}
 \ket{\pFt(s,k)}=\Ut(s,k) \ket{\pFt(0,k)} \:, \\
 \label{eq:Ptrans}
 \PF(s,k;\tau)=\Ut(s,k)\PF(0,k;\tau)\Ut^{\dagger}(s,k) \:,
\end{eqnarray}
from which the second assertion follows.  The third item follows
from proposition \ref{prop:analy}.\qed

$F$ depends on $s,\tau$ and $k$. Because of equations
\eref{eq:Floquet} the $s$ variable is uninteresting. It is a basic
fact of perturbation theory \cite{Kato} that if a (normal)
operator depends analytically on two variables (or more), one can
choose, in general, at most, one variable so that the eigenvectors
and eigenvalues are real analytic. We choose real analyticity in
the $k$ variable. The price one has to pay for this is two fold:
First, the $\tau$ dependence of the projections will, in general, be discontinuous
at
crossings\footnote{This can be seen already for the \msize{2}
matrix function $\tau\sigma_{x} + k\sigma_{z}$ where $\sigma$ are the
Pauli matrices.} and the second is that the $2\pi$ periodicity in
$k$ may be lost. (One can not assume that $\PF$ is both real
analytic and $2\pi$ periodic in $k$.) But,  it is $2\pi N$
periodic for some finite $N$. If there are no crossing  then
$N=1$.

A result of Ferrari  \cite{Ferrari} identifies the Chern numbers of
Floquet eigenstates with the charge they transport and relates the
Chern number of Floquet states to the winding number of the
quasienergies. This identification will play a key
role in the next section where we discuss breakdown.

\begin{prop}\label{thm:FlqChern} Let $\Ft(0,k)$ be a finite
dimensional Floquet operator, which is unitary for real values of
$\tau$ and $k$, and analytic in both. Fix $\tau$ and label the
spectrum so that $\PF_\tau(0,k)$ is real analytic in   $k$ with
period $2\pi N$ . Then
\begin{enumerate}
\renewcommand{\theenumi}{\arabic{enumi}}
\item The quasienergies  $E(k;\tau)$ and the associated
eigenfunctions $\ket{\pFt(s,k)}$ are si\-mul\-tan\-eous\-ly real
analytic in $k$ and $s$.

\item The conductance associated to non-interacting electrons
that fill a k-band of Floquet states is a Chern number multiple of
$1/N$:
\begin{eqnarray}\label{eq:QbyChern}
 Q(\tau;\pFt) =
\frac{1}{2\pi N} \, \imag\,\int_0^{2\pi} \rmd s \,\int_0^{2\pi N}
\rmd k\, \brkt{\frac{\partial \pFt}{\partial s}}{\frac{\partial
\pFt}{\partial k}}\in \frac{\mathbb{Z}}{N} \:.
\end{eqnarray}
\item  $Q(\tau;\pst)$ is also the winding number of the corresponding
eigenvalue,
\begin{equation}\label{eq:QbyWind}
 Q(\tau;\pst)=-\frac{E(2\pi N;\tau)-E(0,\tau)}{2\pi N
} \:.
\end{equation}
\item $ Q(\tau,\pFt)=0$ for all
$\EF>2\max_{s , k \in [0 , 2\pi]}\, \norm{H(s, k)} $.
\end{enumerate}
\end{prop}

\proof\ The first assertion follows from the real analyticity of
the projection, \eref{eq:Psitrans}, and standard facts from
perturbation theory \cite{Kato}. The second assertion follows
from:
\begin{equation}\label{eq:cond}
\tau\,\braket{\pst}{\frac{\partial H}{\partial k}}
{\pst}=\tau\,\frac{\partial}{\partial k}\braket
{\pst}{H}{\pst}+\imag \brkt{\frac{\partial \pst}{\partial k}}
{\frac{\partial \pst}{\partial s}} \:.
\end{equation}
The observable on the left hand side of this identity is the ratio
of the current to the driving electric field (in each $k$
channel). This means $Q$ is a conductance. The identity
\eref{eq:cond} is a consequence of integration by parts, the
equation of motion, and the periodicity in $k$. The quantization
of $Q$ follows by integrating  over the period in $k$ and using
standard facts about Chern numbers. Using the identity
\begin{equation}
\left(\Ft^\dagger \frac{\partial \Ft}{\partial k}\right)(0,k)
=-\rmi\tau\int_0^{2\pi}\, U^\dagger_{\tau} (s,k)\,\left(\frac{
\partial H}{\partial k}\right)\, \Ut(s,k)  \rmd s
\end{equation}
and equation \eref{eq:FlqEigens}
\begin{equation}
\braket{\pFt(0,k)}{\Big(\Ft^\dagger \partial_k \Ft\Big)
(0,k)}{\pFt(0,k)}=\rmi \partial_k E(k;\tau)
\end{equation}
gives assertion 3. Item 4 follows from the observation that the
winding vanishes if $\max\abs{E(k;\tau)}<\pi$. The maximum can
achieved when all the terms in the Dyson expansion of the
evolution operator have the same eigenstate for their maximal
eigenvalue and all the eigenvalues are summed up with the right
signs. It follows that
\begin{equation}
\abs{E} \leq  2\pi\tau \max\norm{H} \:,
\end{equation}
which implies item 4.
\qed

Finally, we note that for the adiabatic limit, $\EF=0$,
eigenstates of the Hamiltonian are also Floquet eigenstates. This
shows that for $\EF=0$ the Chern number for Floquet states is the
Hall conductance of linear response when $H(s,k)$ is the
Hamiltonian for the Hall effect.
\section{Breakdown in Floquet States}\label{sec:breakfloquet}

In the previous section it was shown that the Hall conductance in
Floquet states, for any field strength $\EF$, is related to the
Chern number of the state. Since the Chern numbers are integers,
this implies that the breakdown in Floquet states is always
discontinuous: a quantum phase transition. For $\EF=0$ the
conductance for Floquet states coincide with the Hall conductance
of linear response but when $\EF$ is large enough $Q$ vanish. This
forces a discontinuous breakdown.

At first, it appears that one can argue that a breakdown must
occur for some finite value of $\EF$ because Chern numbers do not
change under small deformations of the bundle of eigenstates, and
so if the Chern number is non-zero at zero field, should it not be
non-zero also for small fields?

To understand how the breakdown occurs we first make the
observation that non-zero Chern numbers for Floquet operators
always come with level crossings:

\begin{thm}\label{thm:WindCross}
Let $\Ft(0,k)$ be a {\em finite dimensional} Floquet operator,
which is unitary for  $\tau,\, k\in \mathbb{R} $, and analytic in
both. Label the spectrum of $\Ft$ so that $\PF_\tau(0,k)$ is real
analytic in $k$ with period $2\pi N$. If any Chern number is
non-zero then $\Ft$ has eigenvalue crossing for some value of $k$.
\end{thm}
\proof\ By general principles, the sum of all the Chern numbers
for any finite dimensional matrix, such as $\Ft$, must vanish.
Hence, if there is a positive Chern number for one of the states
of $\Ft$, there must also be a negative Chern number for some
other state. The quasienergy with a positive winding  must then
cross the one with negative winding.\qed


This leads to a  puzzle that has Chern on one side, and Wigner
von-Neumann (WVN) on the other: Wigner and von-Neumann
\cite{WVN,Knox} say that eigenvalue crossing tend to be unstable,
unless the dimension of parameter space is three or
more\footnote{In the complex case, which is the case relevant
here.}. The relevant parameter space for the Floquet operators is
the $\tau$-$k$ space which has dimension two. This implies that
crossings are unstable. On the other hand Chern numbers give a
topological characterization of the bundle of eigenstates and are
stable under continuous deformations of the bundle. Who wins?

The winners, we claim, are Wigner and von-Neumann. It is, of
course, true that continuous deformation of the bundle keep the
Chern number fixed. But, there is no reason why a continuous
deformations of the Hamiltonian or a variation in $\tau$, should
lead to continuous deformations of the bundle of eigenstates. By
perturbation theory  a small deformation in the Hamiltonian can
result in a discontinuous change of the bundle at points of
crossing. Since non-zero Chern number for Floquet states always
comes with eigenvalue crossing, the bundles always lie at the
boundary of the region of stability.  As a consequence,
generically, at least, Chern numbers for Floquet
operators\footnote{At least, those of the class we study here,
that depend on a single variable $k$} should be zero and the
breakdown should therefore occur at $\EF=0_{+}$.

Experimentally, the breakdown of the Hall effect is continuous and
consequently, the breakdown of Floquet states does not appear to be an
appropriate theory for the breakdown of the Hall effect.
 This  suggests that the breakdown is not a
large time phenomenon. Nevertheless, the breakdown in
Floquet states is an interesting chapter in the theory of
breakdown of Chern numbers.

\section{Unstable Chern numbers}\label{sec:numeric}
We have seen in theorem \ref{thm:WindCross} that non-zero Chern
numbers for Floquet operators always come with eigenvalue
crossings. The question whether Floquet operators do or do not
have stable non-zero Chern numbers, is equivalent to the question
whether Floquet operators do or do not have stable  eigenvalue
winding and crossing.
\subsection{Fragile Winding}

 A simple example, that we
owe to B. Simon,  illustrates how topological objects that are
normally associated with stability become fragile in the context
of eigenvalue problems. Consider the
 unitary
 \begin{eqnarray}
 F_{\tau}(s=0,k)= \left(  \begin{array}{ll}
    \cos(1/\tau)\, e^{ik}   & \sin( 1/\tau) \\
    -\sin( 1/\tau)          & \cos(1/\tau)\, e^{-ik}
\end{array} \right)\:.
\end{eqnarray}
$F$\/ has unit determinant so its two eigenvalues are determined
by a single angle $\phi$. Clearly $\cos\phi=\cos( 1/\tau)\cos k$.
For $1/\tau= \pi n$ the
 two eigenvalues  have
winding numbers $\pm 1$ as $k$ goes through a period.  But, for
most value of $\tau$  the two eigenvalues repel at $0$ and $\pi$,
and have zero winding.  Winding numbers that arise from an
eigenvalue problem are not stable.

\subsection{The Paradigm}

\begin{figure}
  \centering
  \scalebox{0.5}{\includegraphics{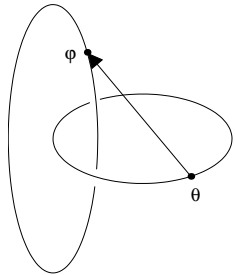}}
  \caption{A map from the 2-torus $(\varphi,\theta)$ to the sphere
  (the unit vector pointing from $\theta$ to $\varphi$). The linking
  number is the degree of the map.}\label{fig:gauss}
\end{figure}

The basic paradigm of line bundles associated with Chern numbers
$\pm 1$ comes from Berry example of spin 1/2 in a magnetic field.
Berry's spin 1/2 model describes a \msize{2} Hamiltonian that is
parameterized by the two sphere.  In the study of Floquet
operators and adiabatic transport, one is interested in families
parameterized by the 2-torus. The question is how to pick a
family of \msize{2} Hamiltonians on a 2-torus with Chern numbers
$\pm 1$. A nice geometric way (\fref{fig:gauss}) to think about
the family is to consider the pull-back of Berry's spin 1/2 on the
2-sphere, by the Gauss map from the 2-torus to the
2-sphere with degree one\footnote{We thank I. Klich for pointing this out.}.

A useful  fact about the quasienergies of \msize{2} Floquet
operators with determinant 1 is that $E_1=-E_2$. Hence, gap
closure occur only at $E=0$ and $E=\pm\pi$. A stable non-zero
Chern number then requires a stable crossing at both at $E=0$ and
$E=\pm\pi$. Since this feature simplifies the numerical analysis
considerably  the \msize{2} Hamiltonian on the torus should also be
traceless.

An example of a periodic, traceless \msize{2} matrix Hamiltonian with Chern
numbers $\pm 1$ is
\begin{eqnarray}\label{eq:H1}
 H_1(s,k)= &\frac{1}
 {\sqrt{\sin (s) ^2 + \abs{a}^2}}\, \left(  \begin{array}{ll}
    \sin(s)            & a \\
    \bar{a} & -\sin(s)
\end{array} \right)
\end{eqnarray}
with
\begin{equation}
a=\cos(k+\frac{\pi}{4})+\rmi\cos(k-\frac{\pi}{4})+(1+\rmi)\left(\cos(s)+1\right)
\:.
\end{equation}

\begin{figure}[htb]
  \centering
  \includegraphics{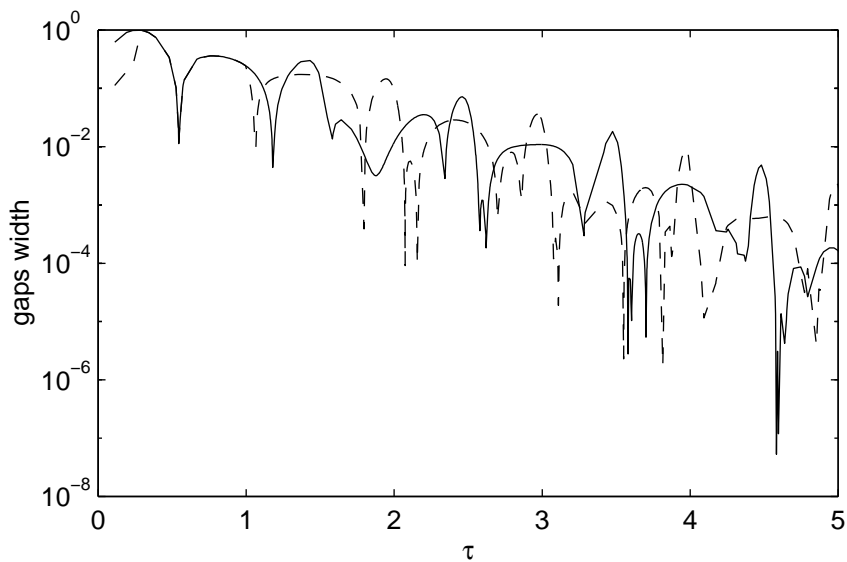}
  \caption{Gaps width of the Floquet operators from $H_1$.
  The solid line is for the gap near $E=\pi$, the dashed line is
  for the gap near $E=0$}
  \label{fig:Gap1}
\end{figure}

The Floquet operator associated to this Hamiltonian has to be
calculated by numerical integration of the \schrod\ equation
\eref{eq:schrodinger}. This Floquet is our basic paradigm.  Now,
if  non-zero Chern numbers for Floquet operators are generically
unstable, this will manifest itself in the opening of gaps at
$\pi$ and $0$. The gaps are shown in figure \fref{fig:Gap1} and
are consistent with the claim that Chern numbers of the Floquet
operator are generically zero.

\subsection{Another Toy Models}

We have also carried out  extensive numerical studies of the
following toy model of a \msize{2}  traceless periodic matrix
family:
\begin{eqnarray}\label{eq:H2eps}
 H_2(s,k,\varepsilon)=
\frac{1}{\sqrt{a^2+(b+\varepsilon)^2+c^2}} \left(
\begin{array}{ll}
   a            & b-\rmi c+\varepsilon \\
  b+\rmi c+\varepsilon & -a
\end{array} \right)  \\
 a= \cos(k) \:,\quad b=\cos(s) \:,\quad c=\cos(s+k) \nonumber \:.
\end{eqnarray}
This Hamiltonian is inspired to some extent by the Harper model on
a triangular lattice \cite{Bellissard}. It has the following
features:
\begin{enumerate}
\item For all $s,k,\varepsilon\neq \pm 1$ the spectrum is constant
$Spec(H_2(s,k,\varepsilon))=\{-1,1\}$.
\item The Chern numbers associated with the eigenvectors of
$H_2(s,k,\varepsilon)$ are $\pm 2$ for  $-1<\varepsilon<1$ and
zero otherwise\footnote{This can be seen by examining sections of
the bundle.}.
\end{enumerate}
We have added an extra parameter, $\varepsilon$ so that  the
associated Floquet operator, $F_\tau(0,k; \varepsilon)$, depends
on three variables, $\{\tau,k,\varepsilon\}$. The reason for
including $\varepsilon$ shall become clear below.

What should one expect for the Floquet operator based on WVN
genericity argument? Since (the unitary) $F_\tau(0,k;
\varepsilon)$ depends on three real variables, one expects
isolated points in $\{\tau,k,\varepsilon\}$ space where there is
crossing. In other words, for some special values of $\varepsilon$
one expects to find crossing, while for most values of
$\varepsilon$, there should be no crossing, and all Chern numbers
should vanish. This turns out to be the case.

Here too the Floquet operator
has to be calculated by numerical integration of the \schrod\
equation \eref{eq:schrodinger}.

\begin{figure}[htb]
  \centering
  \includegraphics{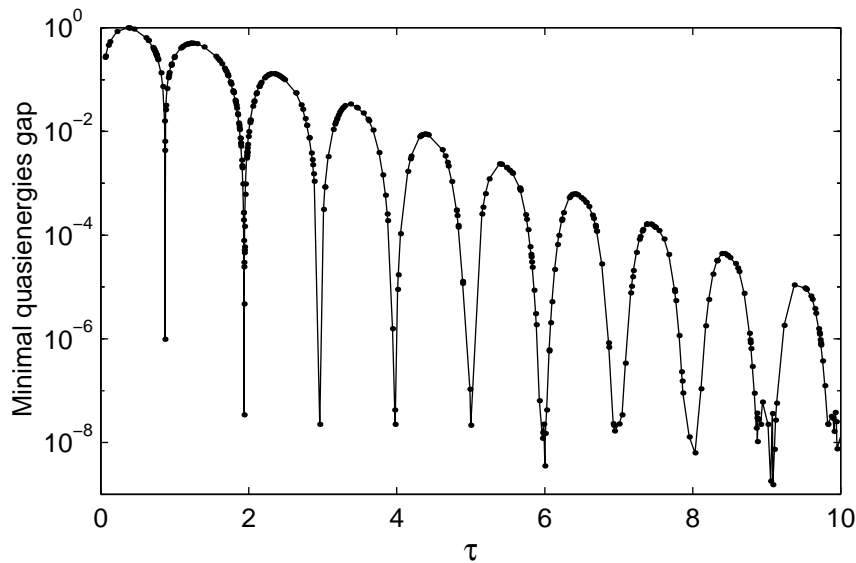}
  \caption{Plot of the minimal quasienergy gap vs.\ $\tau$ for $\varepsilon=0$.
   Notice the
  logarithmic y--scale. The calculation accuracy was of order $10^{-8}$.
  The circles represents the calculation points which are spaced non-uniformly.}
  \label{fig:Gap}
\end{figure}

For $\varepsilon=0$ the crossing at $E=0$ is stable for all
$\tau$. The gap at $E=\pm\pi$ is plotted in \fref{fig:Gap}. As can
be seen, for most values of $\tau$ the gap is wider then the
numerical tolerance. This gap is usually very small. It is easy to
mistake these small gaps for crossing.

Near integers values of $\tau$ the gap width decreases rapidly. At
these points the gaps seem to close, up to our numerical accuracy,
and the winding numbers change from zero to $\pm2$.

These results are in conflict with both the putative stability of
Chern numbers, and the WVN instability of crossing. It is in
conflict with Chern, as non-zero Chern numbers appear to occur at
isolated points on the $\tau$ axis rather than intervals. It is
also in conflict with WVN, for, by the no-crossing rule one
expects no crossing in the $\tau-k$ plane at all. The model would
be consistent with WVN if the point $\varepsilon=0$ turns out to
be a special point.  Although we do not know if and why
$\varepsilon=0$ is special, one expects WVN to fail for {\em some}
$\varepsilon$, and a test would be to see if the crossing
disappear when we wiggle away from $\varepsilon=0$.

\begin{figure}[htb]
  \centering
   \includegraphics{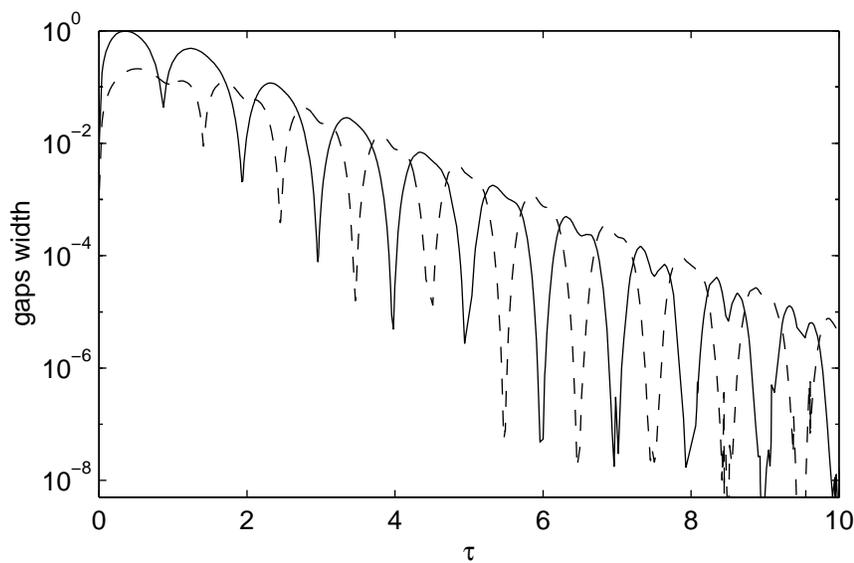}
  \caption{Gaps width of the Floquet operators from $H_2$ with $\varepsilon=0.2$.
  The solid line is for the gap near $E=\pm\pi$, the dashed line is for the gap
near $E=0$}
  \label{fig:Gapeps}
\end{figure}

For non-zero values of $\varepsilon$ both  gaps closure at
$E=0,\pi$ indeed disappear, as is seen in \fref{fig:Gapeps}, and
Wigner and von-Neumann are vindicate.

For more numerical results and description for some of the
numerical procedures see \cite{Thesis}.

In conclusion, the numerical results  for both toy models
  are in agreement with WVN and with our argument that
Chern numbers for Floquet operators are generically zero.

\subsection{The Harper Model}
We have also studied numerically the question of stability of
Chern numbers for the Floquet operator that is associated with the
Harper model.

\begin{figure}[htb]
  \centering
  \includegraphics{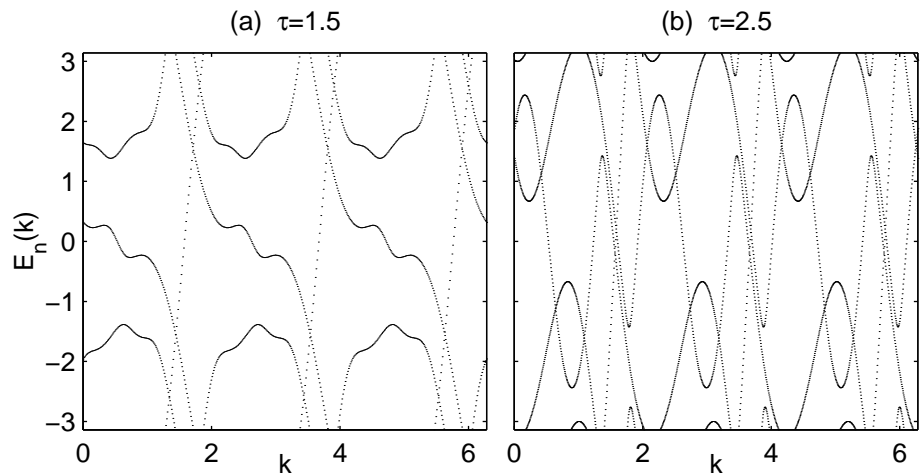}
  \caption{The quasienergies $E(k)$ of the Floquet operator associated
  with the \msize{3} Harper Hamiltonian for $\tau=1.5,2.5$.}
  \label{fig:Bands3}
\end{figure}

\begin{figure}[htb]
  \centering
  \includegraphics{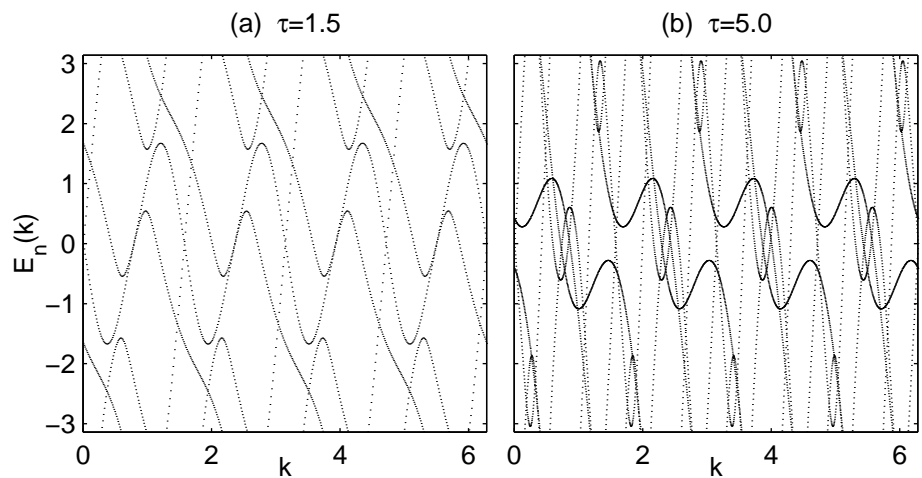}
  \caption{The quasienergies $E(k)$ of the Floquet operator associated
  with the \msize{4} Harper Hamiltonian for $\tau=1.5,5.0$.}
  \label{fig:Bands4}
\end{figure}

Telling a true crossing from an avoided crossing in a numerical
study is often a challenge. Often one finds complicated pictures
where it is difficult to tell with certainty which is really the
case since even real gaps tend to be very small. This is
illustrated by numerical results for the Harper model with $p=1$
and $q=3,4$, see figures \ref{fig:Bands3} and \ref{fig:Bands4}.

Our numerical results are consistent with the assertion that the
Harper model has some stable crossing. It is an open problem to
prove or disprove this.

\section{Failure of Berry's method to detect Floquet crossings}
One reliable method to identify crossing of {\em self-adjoint}
matrices, is to compute a Chern number (in the Hermitian case) or
the Longuet-Higgins phase (in the symmetric case)
\cite{BerryWilk,Berry}. Unfortunately, this method does not work
for the kind of Floquet operators that we consider, as we now proceed to explain.

\begin{figure}[htb]
  \centering
  \includegraphics{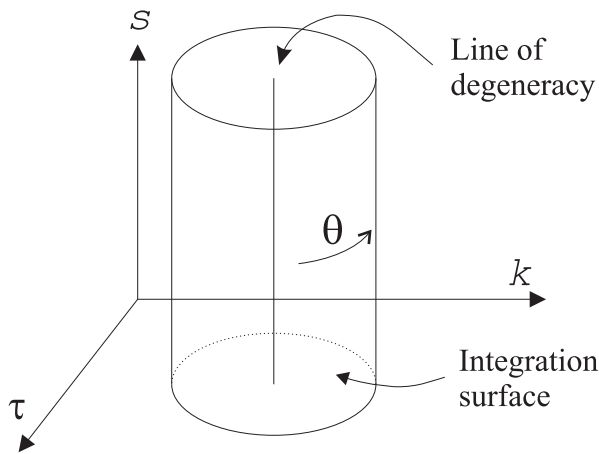}
  \caption{The 2--torus which is used to enclose the crossing line.
Notice that $s$ is a cyclic parameter, so the cylinder closes to a
torus.}\label{fig:Torus}
\end{figure}

Because of the $s$ independence of the spectrum an isolated
crossing would look like the line of constant $k$ and $\tau$ as in
\fref{fig:Torus}. The associated Chern number involves the surface
of integration, a 2-torus of ($s$,$\theta$), as shown in
figure~\ref{fig:Torus}.

Using similar arguments as in equations \eref{eq:QbyChern} and
\eref{eq:QbyWind} we find
\begin{equation}
  \ch(\PF)=-\frac{1}{2\pi} \int_0^{2\pi} \partial_{\theta}
  E(\theta)\,\rmd \theta = -\frac{E(2\pi)-E(0)}{2\pi} \:.
\end{equation}
Since the surface lies away from any crossing
the winding number of all eigenvalues must be
the same. Since the Chern numbers sum to zero, the winding must be
zero too. Chern number are therefore not useful to identify the
crossing of Floquet operators.

\section{Conclusion}

Transport  coefficients {\em in linear response theory}, such as
the Hall conductance, can, in general, undergo a phase transition
as functions of {\em parameters} in the Hamiltonian. This is
because, in the linear regime, the conductance can be expressed in
terms of the spectral data of the Hamiltonian. Spectral data may,
and in general will, lose  smoothness near eigenvalue crossing and
allow for the loss of analyticity.

The breakdown in Chern numbers associated with the Hall
conductance in finite fields seen in \fref{fig:QH3x3} is not
associated with loss of analyticity at finite fields but with an
essential singularity at zero fields.

One way to have Chern numbers undergo a phase transition with the
applied field is to define them by a spectral problem. This is the
case for Chern numbers of Floquet states. An external time
independent electric field, in an appropriate gauge, leads to a
tight-binding Hamiltonians that is periodic in time and admits
Floquet analysis. The Hall conductance in Floquet states reduces
to a spectral problem for the Floquet operator. This holds for any
value of the driving field, and not just in the linear response
regime. Because of this, changing the field can lead to a
non-smooth behavior of the conductance when eigenvalues of the
Floquet operator cross. We argue that the Chern numbers of Floquet
operators are generically zero, because non-zero Chern numbers
 come with eigenvalue crossing. Non-zero Chern numbers
are unstable to small variations in the Hamiltonian and the
breakdown occurs at zero fields.

In conclusion we can say that we find no theoretical reason for
the breakdown of the Hall effect to be a quantum phase transition
at nonzero fields, although we do not rule this out for more
complicated models such as models of infinitely many interacting
particles.  In addition we observe that  under the condition where
one can show that the charge transport is quantized, one can also
show that it is an analytic function of the field with an
essential singularity at zero field.

\ack We acknowledge useful discussions with  A. Auerbach, S.
Fishman, I. Klich, K. Malick, N. Moiseyev, M. Resnikov, B. Simon
and D. Thouless, and especially with A. Elgart. This research was
partially supported by the Israel Science Foundation, SFB 288 and
the fund for the promotion of research at the Technion.

\end{document}